\documentclass[epj]{svjour}
\usepackage{shadow,pifont,epsfig,wrapfig,times,graphics,color}
\usepackage{latexsym}
\usepackage{graphicx, graphics, rotating}
\usepackage{pstcol, pstricks, pst-node, pst-coil, pst-grad}

\begin{document}
\def\dsdt{$\frac{d\sigma}{dt}$}
\def\beqn{\begin{eqnarray}}
\def\eeqn{\end{eqnarray}}
\def\barr{\begin{array}}
\def\earr{\end{array}}
\def\btab{\begin{tabular}}
\def\etab{\end{tabular}}
\def\bite{\begin{itemize}}
\def\eite{\end{itemize}}
\def\bcen{\begin{center}}
\def\ecen{\end{center}}

\def\eq{\begin{equation}}
\def\ee{\end{equation}}
\def\eqa{\begin{eqnarray}}
\def\eea{\end{eqnarray}}
\def\nn{\nonumber}
\def\psmu{P^{\prime \mu}}
\def\psnu{P^{\prime \nu}}
\def\ksmu{K^{\prime \mu}}
\def\pss{P^{\prime \hspace{0.05cm}2}}
\def\psf{P^{\prime \hspace{0.05cm}4}}
\def\kdagger{K\hspace{-0.3cm}/}
\def\ndagger{N\hspace{-0.3cm}/}
\def\psdagger{p'\hspace{-0.28cm}/}
\def\epssdagger{\varepsilon'\hspace{-0.28cm}/}
\def\epsdagger{\varepsilon\hspace{-0.18cm}/}
\def\pdagger{p\hspace{-0.18cm}/}
\def\xidagger{\xi\hspace{-0.18cm}/}
\def\qsdagger{q'\hspace{-0.3cm}/}
\def\qdagger{q\hspace{-0.2cm}/}
\def\keldagger{k\hspace{-0.2cm}/}
\def\ksdagger{k'\hspace{-0.3cm}/}
\def\q2dagger{q_2\hspace{-0.35cm}/\;}
\def\qqs{q\!\cdot\!q'}
\def\lls{l\!\cdot\!l'}
\def\lp{l\!\cdot\!p}
\def\lps{l\!\cdot\!p'}
\def\lsp{l'\!\cdot\!p}
\def\lsps{l'\!\cdot\!p'}
\def\lqs{l\!\cdot\!q'}
\def\pps{p\!\cdot\!p'}
\def\psqs{p'\!\cdot\!q'}
\def\epsp{\varepsilon'\!\cdot\!p}
\def\epsps{\varepsilon'\!\cdot\!p'}
\def\epsl{\varepsilon'\!\cdot\!l}
\def\epsls{\varepsilon'\!\cdot\!l'}
\title{A NEW APPLICATION OF THE G\"URSEY AND RADICATI MASS FORMULA}
\author{M.M. Giannini\inst{1}\and E. Santopinto\inst{1} \and 
A.Vassallo \inst{1}}
\institute{Dipartimento di Fisica, Universit\`a degli Studi di Genova \& 
I.N.F.N. Sezione di Genova}
\date{\today}

\abstract{
We study the spin- and flavour- dependent $SU(6)$ violations in the baryon 
spectrum by means of a G\"ursey Radicati mass formula.
The average energy of each $SU(6)$-multiplet is described using the
$SU(6)$ invariant interaction given by a hypercentral potential containing 
a linear and a hypercoulomb term. We show that the non strange and strange 
baryon masses are in general fairly well reproduced and moreover that 
the G\"ursey  Radicati formula holds in a satisfactory way also for the 
excited states up to 2 GeV. The coefficients of the 
G\"ursey Radicati $SU(6)$ breaking part obtained by the fit of 
the three-quark spectrum can be used to evaluate in first approximation the
splitting within multiplets also for exotic baryon systems. 
}
\maketitle
\section{Introduction}
\noindent Different versions of  Constituent Quark Models (CQM) have been
proposed in the last decades in
order to describe the baryon properties.  What they have in common 
is the fact that the three-quark interaction
can be separated in two parts: the first one, containing the confinement
interaction, is spin and flavour independent and
it is therefore
$SU(6)$ invariant, while the second violates the $SU(6)$ symmetry. This
separation has been supported by the very first
Lattice QCD calculations
~\cite{deru} and is confirmed by the most recent ones~\cite{bali12,alexandrou}.
The various CQMs differ in the way the $SU(6)$
invariance is violated. One of the most popular ways was the introduction
of a hyperfine (spin-spin) interaction
~\cite{ik,gi,pl,ff}, however in many studies a spin- and isospin-
~\cite{bil,olof,ple,iso} or a spin- and flavour-dependent interaction
~\cite{bil,olof,ple} has been considered.
In this paper we study the symmetries of the baryon spectrum using  
a very simple approach based on the G\"ursey Radicati mass formula (GR)
~\cite{GR}. It is well known that the baryon spectrum exhibits an approximated 
$SU(6)$ symmetry and that the GR mass formula, despite it's simplicity, 
describes quite well the way this symmetry is broken, at least in the lower 
part of the baryon spectrum.  
Our idea is to build up a very simple model based on the GR mass formula,
to fix the parameters of the model in way to obtain the best description 
(within the limits of this approach) of the baryon spectrum 
and thereafter use the model (with the parameters values fixed on the baryon 
spectrum) to give predictions for the masses of 
other hadronic systems as for example pentaquarks~\cite{bgs}.
The model we propose is a simple CQM where the $SU(6)$ invariant part of 
the Hamiltonian is the same as in the Hypercentral Constituent Quark 
Model~\cite{pl,ff} and where the $SU(6)$ symmetry is broken by a GR inspired 
interaction.\\   
In the second section we briefly remind the hypercentral CQM and its
results, then in the third section we construct the model using the
GR as the $SU(6)$ breaking term. In the fourth section we give the results 
obtained by fitting the GR parameters to the strange and non strange baryon 
energies and we compare the spectrum with the experimental data. 
Finally we discuss our results.

\section{The hypercentral model} 
\label{sec:1} 
\noindent 
The experimental $4-$ and $3-$star non strange resonances can be arranged
in  $SU(6)$ multiplets. This means that the quark dynamics
has a dominant $SU(6)$ invariant part, which accounts for the average
multiplet energies. In the Hypercentral Constituent Quark Model (hCQM) 
it is assumed to be given by the hypercentral potential~\cite{pl}
\begin{equation}\label{eq:pot}
V(x)= -\frac{\tau}{x}~+~\alpha x,
\end{equation}
where 
\begin{equation}
x=\sqrt{{\mbox{\boldmath $\rho$}}^2+{\mbox{\boldmath $\lambda$}}^2},
\end{equation}
is the hyperradius and 
$\mbox{\boldmath $\rho$}$ and $\mbox{\boldmath
$\lambda$}$ are the Jacobi coordinates describing the internal quark motion.
Interactions of the type linear plus Coulomb-like have been used since
long time for the meson sector, {\em e.g.} the Cornell potential. 
This form has also 
been obtained in recent Lattice QCD calculations~\cite{bali12,alexandrou} for
$SU(3)$ invariant static quark sources.
Introducing, along with the hypercentral potential, a standard 
hyperfine interaction~\cite{ik} which breaks the $SU(6)$ symmetry, the hCQM 
has given a fair description of the non strange baryon spectrum~\cite{pl} and 
of other various physical quantities of interest: 
the photocouplings~\cite{aie}, the electromagnetic transition amplitudes 
~\cite{aie2}, the elastic nucleon form factors~\cite{mds} and the ratio 
between the electric and magnetic proton form factors~\cite{rap}.\\
Subsequently, in order to improve the description of the non strange spectrum, 
an isospin dependent $SU(6)$ violating term has been introduced~\cite{iso}. 
The complete interaction used in this latter case is given by 
\begin{equation}\label{eq:tot}
H_{int}~=~V(x) +H_{\mathrm{S}} +H_{\mathrm{I}} +H_{\mathrm{SI}}~,
\end{equation}
where $V(x)$ is the linear plus hypercoulomb SU(6)-invariant potential of
eq.(\ref{eq:pot}), while $H_{\mathrm{S}}$ is the hyperfine interaction and
$H_{\mathrm{I}}$, $H_{\mathrm{SI}}$ are respectively isospin and spin-isospin 
dependent terms. 
Similar results can be obtained in a relativized version of the model
~\cite{rel}, in which the quark kinetic energy has the correct relativistic 
form.\\
The preceding results show that both spin and isospin (or flavour)
dependent terms in the quark
Hamiltonian are important. Their contributions can be considered as
perturbative $SU(6)$-violating
terms added to the unperturbed $SU(6)$-invariant energies provided by the
hypercentral potential of eq. (\ref{eq:pot}).

\section{The strange baryon spectrum and the G\"ursey Radicati mass
formula}

\noindent
The spin and isospin dependent interactions considered in the previous
section are not the only
source of $SU(6)$ violation. In order to study the strange baryon
spectrum one has to consider the $SU(3)$ violation produced by the differences 
in the quark masses. The well known Gell-Mann-Okubo (GMO) mass formula 
~\cite{gmo} made use of a  $\lambda_8$ violation of $SU(3)$ in order to 
describe the mass splittings  within the various $SU(3)$ multiplets; 
according to this formula the mass $M$ of a baryon belonging to a given 
$SU(3)$ multiplet can be expressed as
\begin{equation}
\label{eq:gmo}
M=M_0+D~Y+E~\left [T(T+1)-\frac{1}{4}Y^2 \right ],
\end{equation}
\noindent where $M_0$ is the average energy value of the $SU(3)$ multiplet, 
$Y$ is the hypercharge, $T$ is the Isospin of the baryon and $D$ and $E$ are 
parameters to be fitted.
A simple way to interpret the origin of such a violation is just to attribute 
to the strange quark a mass different from the up and down quark ones. 
The calculations were performed without reference to any
explicit dynamical model, but using standard group theoretical methods. 
The unknown parameters $D$ and $E$ in the $SU(3)$ violating terms can be in
principle fitted to the experimental masses, thereby providing a
phenomenological way to describe the spectrum.\\
A similar approach for the description of the splittings within the
SU(6) baryon multiplets is provided by the G\"ursey Radicati mass formula 
~\cite{GR}. In the original paper the mass formula reads:
\begin{equation}
\label{eq:groriginal}
M=M_0+CS(S+1)+DY+E\left [T(T+1)-\frac{1}{4}Y^2 \right ]
\end{equation}
\noindent where $S$ is the spin. Eq.(\ref{eq:groriginal}) can be rewritten 
in terms of Casimir operators in the following way
\begin{eqnarray}
  \label{eq:grorigcasimir}
  M=M_0+C C_2[SU_S(2)]+D C_1[U_Y(1)]\nonumber \\
        +E\left [C_2[SU_I(2)]-\frac{1}{4}(C_1[U_Y(1)])^2 \right]
\end{eqnarray}
\noindent where $C_2[SU_S(2)]$ and $C_2[SU_I(2)]$ are the $SU(2)$
(quadratic) Casimir operators for
spin and isospin, respectively, and $C_1[U_Y(1)]$ is the Casimir for the
$U(1)$ subgroup generated by the hypercharge $Y$. 
The presence of a spin dependent term is necessary since states belonging to
a definite $SU(6)$ multiplet do not have the same spin value. This mass formula
has proven to be successful in the description of the gruond state baryon 
masses, however, as stated by the authors themselves,
eq. (\ref{eq:grorigcasimir}) is not the most general mass formula that
can be written on the basis of a broken $SU(6)$ symmetry.\\
In order to generalize eq. (\ref{eq:grorigcasimir}) one can consider a
dynamical spin-flavour symmetry
$SU_{SF}(6)$ and write the following chain of subgroups
\begin{equation}
\label{eq:chain}
\begin{array}{cccccccccccc}
SU_{SF}(6) & \supset & SU_F(3) & \otimes & SU_S(2)& \supset & SU_I(2) &
\otimes
& U_Y(1) & \otimes & SO_S(2) &   \\
\downarrow &  & \downarrow &  & \downarrow &  & \downarrow & & \downarrow
& &
\downarrow &  \\
(\lambda_1,..\lambda_5) &  & (\lambda_f,\mu_f) & &
S & & I & & Y & & M_S&
\end{array}  
\end{equation}
where in the bottom row we report the quantum numbers which label the
irreducible representations of the corresponding groups.
Therefore one can describe the $SU_{SF}(6)$ symmetry breaking mechanism
by generalizing eq. (\ref{eq:grorigcasimir}) as
\begin{eqnarray}
  \label{eq:grfull}
  M=M_0+AC_2[SU_{SF}(6)]+BC_2[SU_F(3)]\nonumber \\
       +CC_2[SU_S(2)]+DC_1[U_Y(1)] \nonumber \\
       +E\left(C_2[SU_I(2)]-\frac{1}{4}(C_1[U_Y(1)])^2\right)
\end{eqnarray}
The generalized G\"ursey Radicati mass formula eq. (\ref{eq:grfull}) can
be used to describe the whole baryon spectrum, provided that two conditions 
are fulfilled. The first condition is the possibility of
using the same splitting coefficients for different $SU(6)$ multiplets.
This seems actually to be the case, as shown by the algebraic approach to 
the baryon spectrum~\cite{bil}, where a formula similar
to eq. (\ref{eq:grfull}) has been applied. The second condition is given
by the possibility of getting reliable values for the unperturbed mass 
values $M_0$.
Our idea is to use for this purpose the $SU(6)$ invariant part of the hCQM,
which provides a good description of the 
non strange baryon spectrum and to introduce a G\"ursey Radicati inspired 
$SU(6)$ breaking interaction to describe the splittings within each $SU(6)$ 
multiplet. We shall therefore make use of the following three quark Hamiltonian
\begin{equation}
  \label{eq:hcqmhamilt}
  H=H_0+H_{GR}
\end{equation}
with
$$H_0=3m+\frac{\mbox{\boldmath $p$}_{\lambda}^2}{2m}+\frac{
\mbox{\boldmath $p$}_\rho^2}{2m}+V(x)~,$$
and
$$H_{GR}=+A÷C_2[SU_{SF}(6)]+B÷C_2[SU_F(3)]+C÷C_2[SU_S(2)]$$
$$+DC_1[U_Y(1)]+E÷\left(C_2[SU_I(2)]-\frac{1}{4}(C_1[U_Y(1)])^2\right)~,$$
where $V(x)$ is the hypercentral potential of eq.(\ref{eq:pot}), 
and  m is the constituent quark mass. 
It has to be remarked that, in order to simplify the solving procedure, 
the constituent quark masses are assumed to be the same for all the quark 
flavours ($m_u=m_d=m_s=m$), therefore, within this approximation, the 
$SU(3)$ symmetry is only broken dynamically by the spin and flavour dependent 
terms in the Hamiltonian. In other words, in this approximation, 
the effects of the strange quark mass on the baryon 
spectrum are described by the two terms of eq.(\ref{eq:gmo}). \\
The eigenproblem of $H_0$ can be solved numerically,
the spin-flavour part of the resulting eigenfunctions has definite
properties under transformations of the $SU_{SF}(6)$ group and its subgroups. 
Using the notation of equation (\ref{eq:chain}) the spin-flavour part of the
wave function can be written as
\begin{equation}
\label{eq:spinflavourpart}
\left \vert
(\lambda_1,\lambda_2,\lambda_3,\lambda_4,\lambda_5),(\lambda_f,\mu_f),
I,Y,S,M_S \right \rangle~.
\end{equation}
Often the irreducible representations are identified not by the
quantum numbers but by
their dimension. Thus, for example
$$\left \vert (3,0,0,0,0), (2,1), I=\frac{1}{2}, Y=1, S=\frac{1}{2},M_S
\right \rangle$$
$$\equiv \vert ^{2}8_{1/2},[56,0^{+}],N \rangle $$
is the spin-flavour part of the nucleon wave function. The notation used
is
$$\vert^{2S+1}\mbox{dim}(SU(3))_{J},[\mbox{dim}(SU(6)),L^{P}], X \rangle,$$
where dim$(SU(n))$ is the dimension of the $SU(n)$
representation, $S$ and $L$ are the total spin and orbital angular
momentum of the quark system,
respectively, $J$ and $P$ are the spin and parity of the resonance and
$X = N,\Delta, \ldots$ denotes the type of baryon resonance.\\
The action of $H_{GR}$ on the eigenstates of $H_0$  is completely 
identified by the expectation values of the Casimir operators on
the states of eq. (\ref{eq:spinflavourpart})
\begin{eqnarray}
  \label{eq:casimir}
  \langle C_2[SU_{SF}(6)] \rangle & = & \left \{
  \begin{array}{ccc}
  45/4 & \mbox{for} & [56] \\
  33/4 & \mbox{for} & [70] \\
  21/4 & \mbox{for} & [20]
  \end{array}  \right . \nonumber \\
 \langle C_2[SU_{F}(3)] \rangle & = & \left \{
  \begin{array}{ccc}
  3 & \mbox{for} & [8] \\
  6 & \mbox{for} & [10] \\
  0 & \mbox{for} & [1]
  \end{array}  \right . \nonumber \\
\langle C_2[SU_I(2)] \rangle & = & I(I+1) \nonumber \\
\langle C_1[U_Y(1)] \rangle & = & Y \nonumber \\
\langle C_2[SU_S(2)] \rangle & = & S(S+1)
\end{eqnarray}
Therefore the mass of each baryon state $\vert B \rangle$ can be written
as:
\begin{equation}
  \label{eq:masses}
  \langle B \vert H \vert B \rangle =E_{\gamma\nu}+ \langle B \vert
H_{GR}
\vert B \rangle
\end{equation}
where $E_{\gamma\nu}$ denotes the eigenvalue of $H_0$ with
$\gamma=2n+l_{\rho}+l_{\lambda}$ ($n$ being a non negative integer), 
$\nu$ denotes the number of nodes of the space three quark wave functions 
and $l_{\rho}$, $l_{\lambda}$ are the orbital 
angular momenta corresponding to the Jacobi coordinates (see e.g.~\cite{pl}).\\
Since $H_{G.R.}$ does not depend on the spatial degrees of freedom, 
the $SU(6)$ 
breaking term introduced in this model is diagonal in the baryon states, this 
means that the G\"ursey Radicati term is able to give energy 
splittings within the $SU(6)$ multiplets, but no configuration mixing effects 
can arise from such an interaction
\footnote{The kind of problems that one can face neglecting the 
spatial dependence on the $SU(6)$ breaking part is discussed 
by Jennings and Maltman \cite{Jennings:2003wz}}. 
Therefore the model is expected to fail in the description of all thoose 
observables where a good description of the three quark wave function is 
crucial.  

\section{Results}

\begin{figure*}[!ht]
  \centering
  \resizebox{0.7\textwidth}{!}{%
  \includegraphics{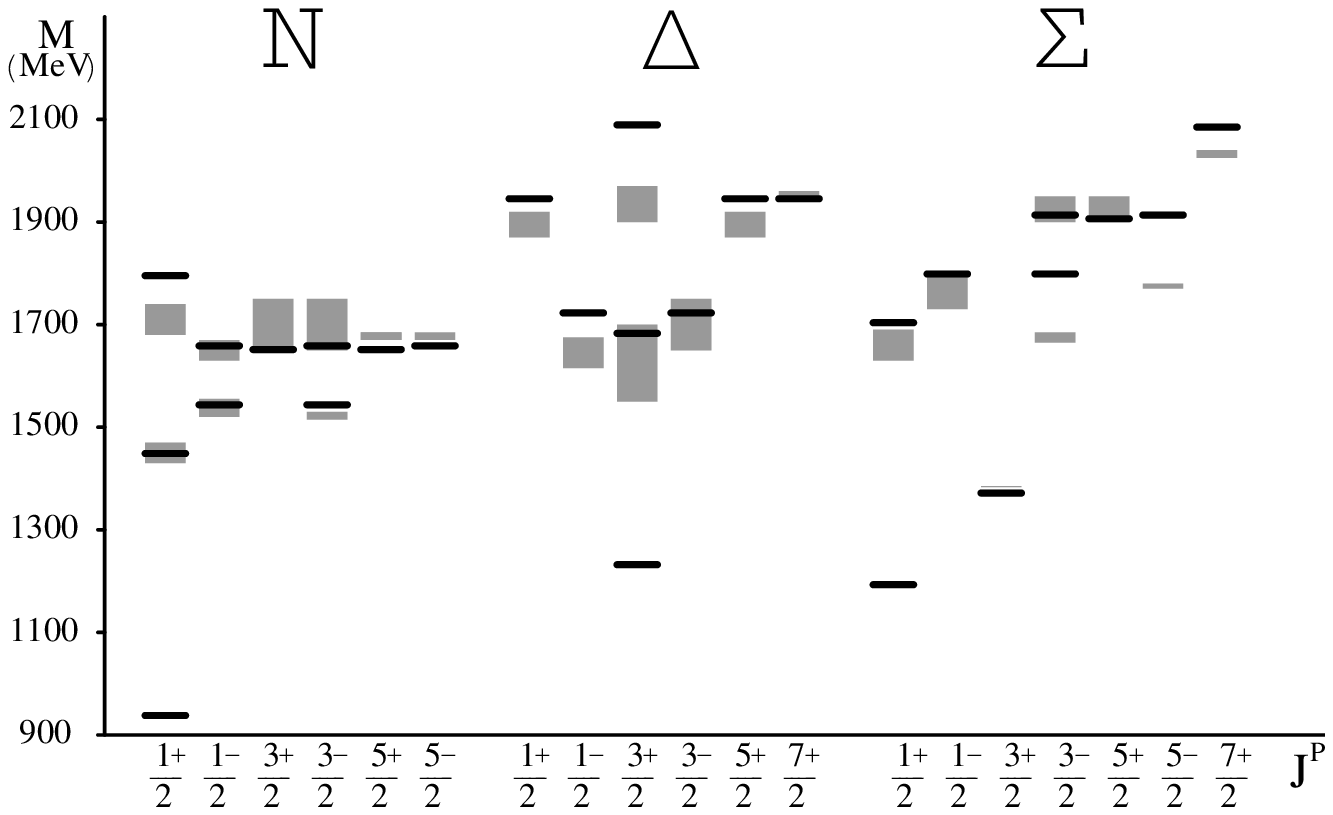}
} \\
\vspace*{.5truecm}
\resizebox{0.7\textwidth}{!}{%
  \includegraphics{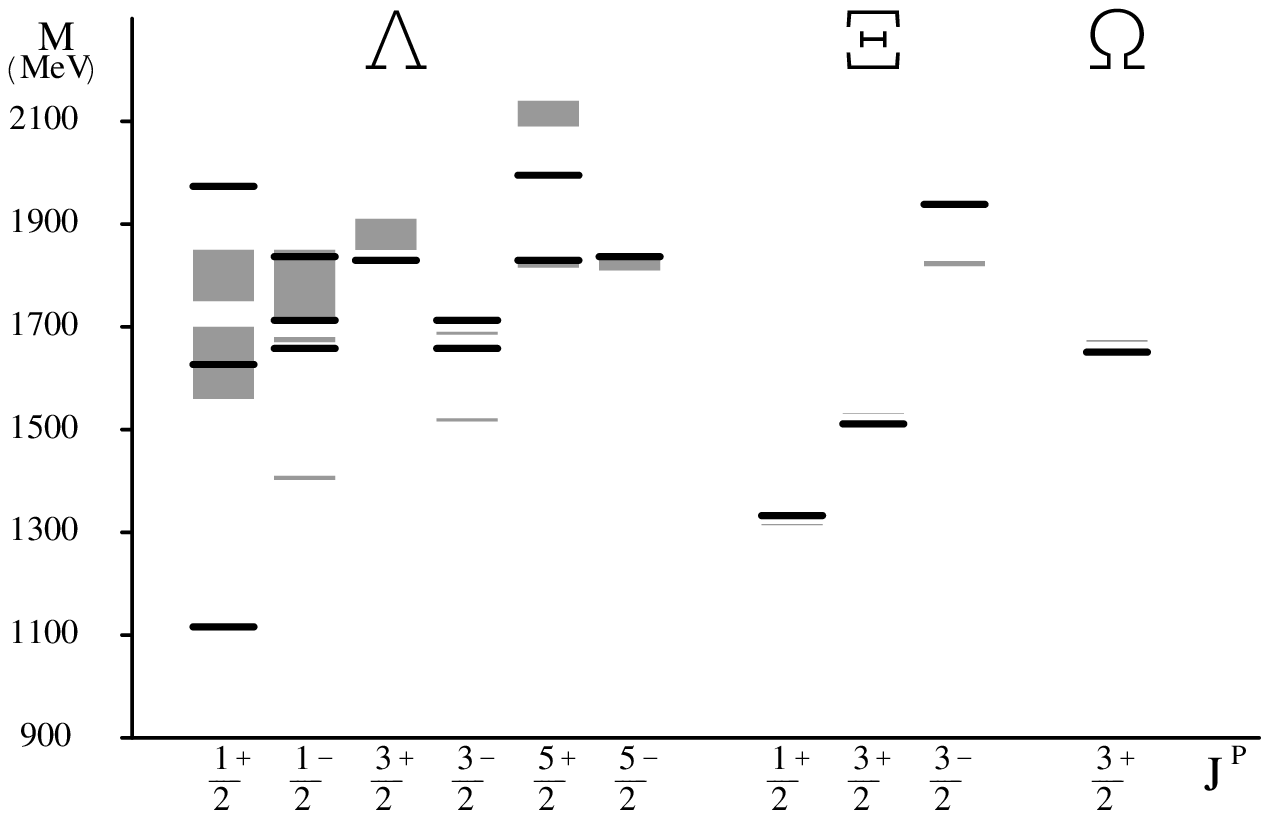}
} 
  \caption{The energy levels (black lines) for the 3 and 4 star resonances 
obtained with the Hamiltonian (\ref{eq:hcqmhamilt})
fixing the parameters as described in equation (\ref{eq:fixingparameters}). 
The numerical values of the calculated masses of baryon 
resonances are reported in Table \ref{tab:tabstrani1}, column
$M^{I}_{calc}$.  The  values of the parameters of the Hamiltonian are reported 
in column (I) of Table \ref{tab:paramgr1}. 
The experimental data are taken from PDG~\cite{pdg} (gray boxes).}
  \label{fig:strani1}
\end{figure*}

\begin{figure*}[!ht]
  \centering
  \resizebox{0.7\textwidth}{!}{%
  \includegraphics{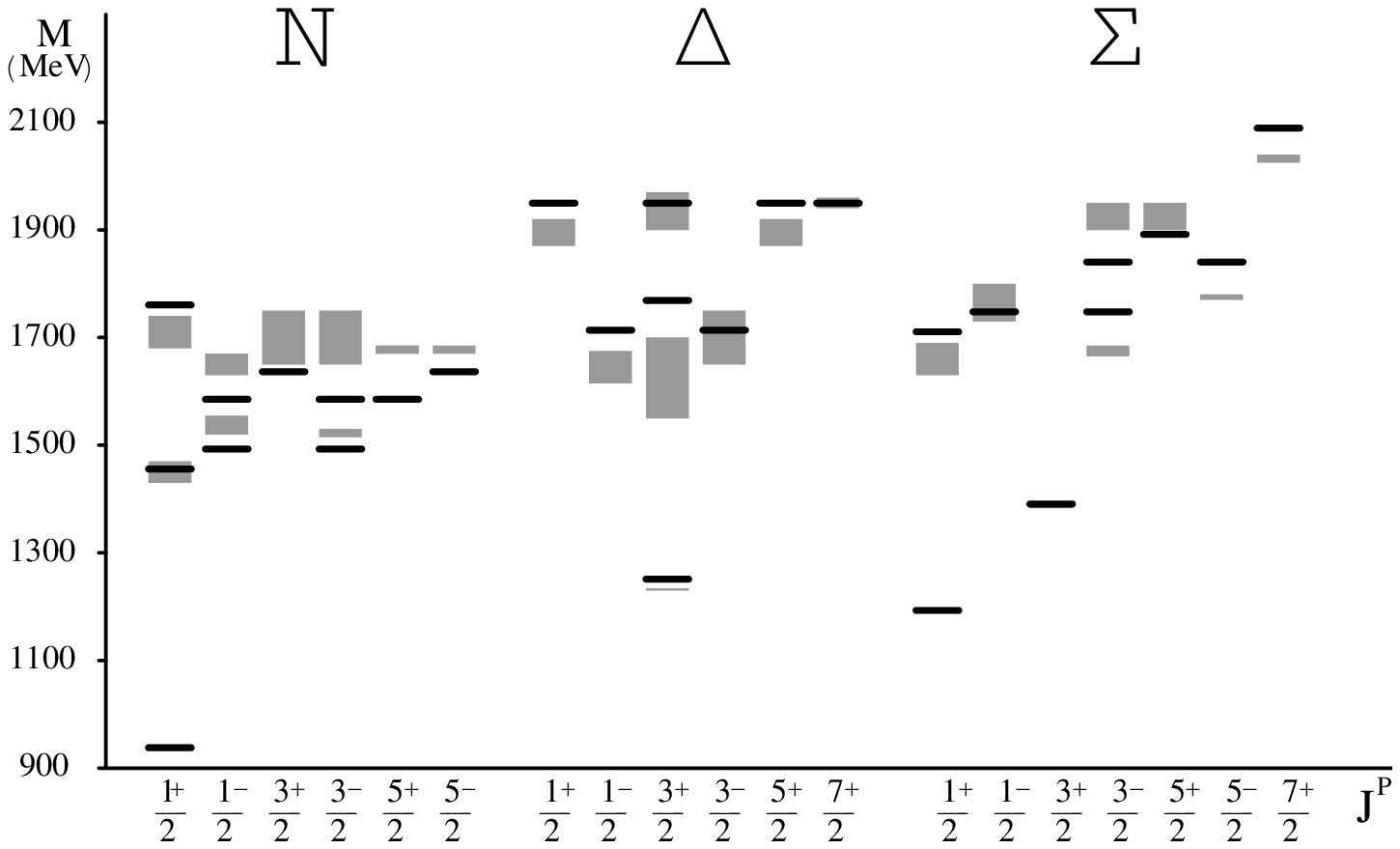}
} \\
\vspace*{.5truecm}
\resizebox{0.7\textwidth}{!}{%
  \includegraphics{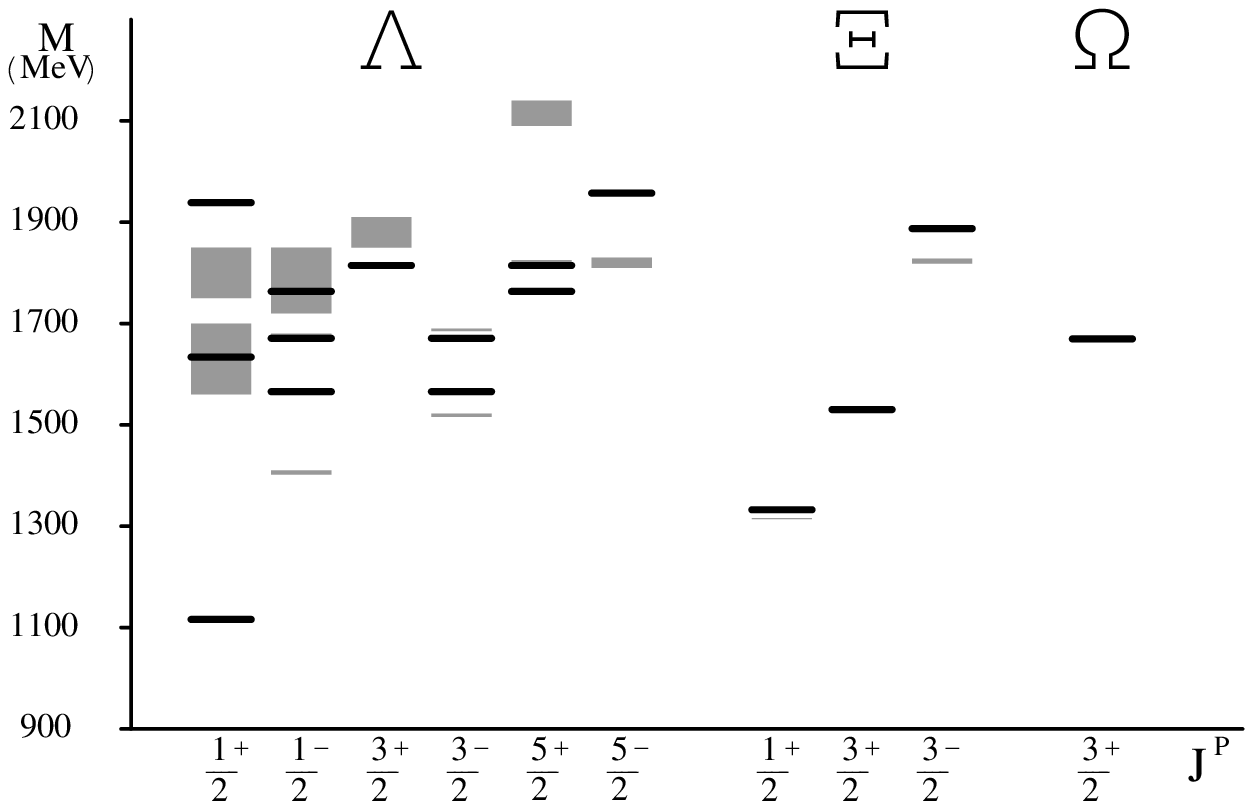}
} 
  \caption{The energy levels (black lines) for the 3 and 4 star resonances 
obtained with the Hamiltonian (\ref{eq:hcqmhamilt}) fixing the Hamiltonian 
parameters by a fitting procedure. The resulting
values of the parameters are reported in column II  of Table
\ref{tab:paramgr1}, while the
numerical values of  the calculated masses are also reported in Table
\ref{tab:tabstrani1}, column
$M^{II}_{calc}$. The experimental data are taken from PDG~\cite{pdg}
(gray boxes).}
  \label{fig:strani2}
\end{figure*}

\noindent 
The parameters in $H_{GR}$ can be determined in order to reproduce the
experimental values of the energy splittings.
We first adopt an analytical procedure by means of which we choose a
limited number of well known
resonances and express their mass differences using $H_{GR}$ and the
Casimir operator expectation values given
in the previous section. We list in the following the analytical
expressions for the mass differences
of the chosen pairs of resonances:
\begin{eqnarray}
  \label{eq:fixingparameters}
  (N(1650)S11-N(1535)S11)&=&3C
\nonumber \\
  (\Delta(1232)P33-N(938)P11)&=&9B+3C+3E
                                           \nonumber \\
  (N(1535)S11-N(1440)P11)&=&(E_{10}-E_{01})+12A
\nonumber \\
  (\Sigma(1193)P11-N(938)P11)&=&\frac{3}{2}~E-D
\nonumber \\
  (\Lambda(1116)P01-N(938)P11)&=&-D-\frac{1}{2}~E .
\end{eqnarray}
Looking at eq. (\ref{eq:fixingparameters}) it is easy to understand that
for
the description of the non-strange baryon spectrum the only relevant
parameters are $A$, $C$ and the combination $(3B+E)$. It should be noted
that, in order to apply
the  G\"ursey Radicati mass formula to the excited states, it is necessary
to know the coefficient $A$
of the $SU(6)$ Casimir operator and the excited energies provided by the
CQM.\\
Once the SU(6) breaking interaction has been determined, the parameters of
$H_0$ ($\alpha$ and $\tau$) which lead to the unperturbed energies 
$E_{\gamma \nu}$
can be fixed by a minimization procedure on the non-strange baryon
spectrum. The complete list of the
parameter values is reported in Table \ref{tab:paramgr1}, column (I). In
this way the
$E_{\gamma \nu}$ levels are placed close to the  central mass value of
each SU(6) multiplet. As shown in eq.
(\ref{eq:fixingparameters}), a further adjustment to the unperturbed multiplet 
energy is provided by
the presence of the $SU(6)$ Casimir operator. The resulting spectrum is
shown in  Fig. \ref{fig:strani1} and
Table \ref{tab:tabstrani1} column $M^{I}_{calc}$.
Despite of the simple form of the SU(6) breaking interaction, the general
features of the
spectrum are fairly well reproduced, especially in the low energy part. It
has to be noted in particular
that, besides the ground state masses which have been fixed through
eq.(\ref{eq:fixingparameters}), the
predicted  masses of the $\Sigma^*$, $\Xi$, $\Xi^*$ and $\Omega$ states
are nicely close to the
experimental values.\\
The second approach followed in the application of the G\"ursey Radicati
mass formula is to fit all
parameters at the same time in order to obtain the best reproduction of
the spectrum of the 3 and 4 star resonances
\footnote{The PDG~\cite{pdg} quotes a three stars $\Xi(1690)$ resonance; 
however, since the values of spin and parity are not 
known, this resonance cannot be identified with a definite eigenstate of the 
Hamiltonian and therefore this state has been excluded from our analysis.}.
The fitted parameters are reported in Table
\ref{tab:paramgr1}, column (II), while the resulting
spectrum is shown in Fig. (\ref{fig:strani2}) and the corresponding
numerical values are given in Table
\ref{tab:tabstrani1}, column $M^{II}_{calc}$. The result is a better
overall agreement with the
experimental data, even if the prediction in the non strange sector is
worsened. For this reason, we prefer
the values of the parameters obtained with the previous analytical method
since we have used only well known
and well established resonances in order to fix the parameters.\\
In both cases a non zero value of the $SU(6)$ Casimir coefficient is
needed in order to reproduce the average multiplet
energies. We have also tried a fit imposing $A=0$. The resulting
parameters $\alpha$ and $\tau$ are however considerably
different with respect to those of Table \ref{tab:tabstrani1} because the
lack of the parameter $A$
must be compensated by the $SU(6)$ invariant energies provided by the
hypercentral potential.
This is particularly evident in the case of the negative parity 
resonances, where the energy difference
$E_{10}-E_{01}$ must be bigger than in the previous case in order to
obtain a good reproduction of the masses; in this way, however, 
the right ordering of the Roper resonance and the negative parity
resonances is lost. This means that the presence of the Casimir
$C_2[SU_{SF}(6)]$ is needed and its effect is to shift down the energy of
the first excited $0^+$ state with respect to
that of the $1^-$. The effect of this term is very similar to that of the
phenomenological $U$ potential of the
Isgur-Karl model \cite{ik}\\
The mass formula of eq.(\ref{eq:grfull}) can be used to add a simple
SU(6) breaking interaction to a CQM
and despite its simplicity it gives rise to a good description of the
baryon spectrum. Of course for the wavefunctions,
and other observables, it is not expected to be as successful as for the
spectrum.\\
Another important feature of this kind of approach is the model
independence of the $SU(6)$ breaking
part of the Hamiltonian. Looking at eq.(\ref{eq:fixingparameters}) it is
easy to understand that the values of the
parameters of the SU(6) breaking part of the Hamiltonian ({\em i.e.} the
B,C,D,E parameters) are completely independent
on the choice of $H_0+AC_2[SU_{SF}(6)]$ which must describe the  central
mass value of each $SU(6)$ multiplet.
An important consequence of the independence, in first approximation, of
the coefficients of the Casimir operators
on the particular wave functions, is the possibility of using the same
$SU(6)$ breaking Hamiltonian for systems with a
different number of quarks. This has been done in a recent study of the
pentaquark spectrum, where the G\"ursey Radicati
mass formula of eq.(\ref{eq:grfull}) has been used for a systematic
analysis of the ($SU(6)$ breaking) splittings within
the exotic baryon multiplets~\cite{bgs}.\\
Finally, we present some comments on the G\"ursey Radicati mass formula of
eq.(\ref{eq:grfull}). As
we have already observed, the last two terms, that is those with
coefficients $D$ and $E$, describe up to first order the $SU(6)$-violation
coming from the mass
difference of quarks, as it has been done in the Gell-Mann-Okubo formula.
The remaining terms are
expected to appear once an explicit dynamics for the quark system is
introduced. For example, in a
recent calculation of multiquark state energies, a spin-flavour dependent
interaction of the type
$$H_{\lambda \sigma}=
\sum^{n}_{i<j} (\mbox{\boldmath $\lambda$}_i\cdot\mbox{\boldmath
$\lambda$}_j)(\mbox{\boldmath $\sigma$}_i\cdot\mbox{\boldmath $\sigma$}_j)$$ 
has been introduced~\cite{helminen};
${\boldmath\lambda}_{i}$  are the $SU_{flavour}(3)$
matrices. The
matrix elements of such spin-flavour interaction between states belonging
to  definite irreducible
representations of $SU(6)$, $SU(3)_{flavour}$ and $SU(2)_{spin}$   can be
calculated as
\begin{eqnarray}
\label{eq:spinflavour}
\left \langle [f]^{SU(6)}[f]^{SU(3)}[f]^{SU(2)}\right \vert H_{\lambda \sigma}
\left \vert [f]^{SU(6)}[f]^{SU(3)}[f]^{SU(2)}\right \rangle
\nonumber \\
 =4C_2(SU(6))-2C_2(SU(3))-\frac{4}{3}C_2(SU(2))-8N_q\hspace*{0.5truecm},
\end{eqnarray}
\noindent where $N_q$ is the number of quarks~\cite{helminen} and
$C_2(SU(2))$ is given by
$S(S+1)$,  $S$ being the total spin. 
If the spatial dependence of the $SU(6)$ breaking
terms is not neglected, this is no more true.\\  
As a conclusion, we can say that the G\"ursey Radicati (\ref{eq:grfull}) 
is a simple way to parametrize at the first order the possible
$SU(6)$-breaking terms of the strong interaction. 
The approach we have adopted here is then to parametrize all the
$SU(6)$-breaking terms by means of
the generalized G\"ursey Radicati, without formulating any hypothesis on
the microscopic
mechanism (one-gluon exchange, Goldstone-boson interaction, chiral
soliton ...). The next step will
be to introduce an explicit $SU(6)$ breaking term, containing also the
spatial dependence and with a clear microscopic origin.

\section{Discussion}

\noindent
We have shown that the G\"ursey Radicati (GR) mass formula is a good
parametrization
of the baryon energy splittings coming from $SU(6)$ breaking. The
splittings are
considered as perturbations superimposed to the $SU(6)$ invariant levels,
which, in
our approach, are given by the hypercentral three-quark potential
~\cite{pl}. The overall good
description of the spectrum which we obtain shows that the GR mass
formula can also be used to give a fair description of the energies 
of the excited multiplets at least up to 2 GeV and not only for the
ground state octets and decuplets, where it has been originally applied. 
There are still problems with the reproduction of some hyperons, 
in particular for the  $\Lambda(1405)$ and the 
$\Lambda(1520)$ resonances that come out degenerate and above the experimental 
values.
There are problems in the reproduction of the experimental masses also in 
the $\Sigma$ sector
where both the $\Sigma(1670)D13$ and the $\Sigma(1775)D15$ four stars 
resonances turn out 
to have predicted masses about $100$ MeV above the experimental values.  
A better agreement can be obtained  either using the square of the 
mass~\cite{bil} or trying to
include a
spatial dependence in the $SU(6)$-breaking part, which may have, among
others, a
delta or Gaussian factor, in order to decrease the breaking with the
increase of the
spatial excitation. Although the space dependence of the $SU(6)$ breaking
terms has
to be neglected in order to apply the GR formula, We can consider the 
G\"ursey Radicati $SU(6)$ breaking as
the first
order parametrization of the splittings due to an interaction which
depends also on
the coordinates. Within this approximation it can be used for the
description of the
SU(6)-breaking effects independently from the way in which one describes
the spatial
part. It can be applied, with the same coefficients, also to systems with
different
number of quarks, such as baryons or pentaquarks. A similar statement is
valid if we
restrict ourselves to $SU(3)$ breaking, using a Gell-Mann-Okubo mass
formula. In
fact, the representations involved will differ for each system and the
dependence on
the number of quarks will be accounted for by the different values of the
various
Casimir operators. 
On the contrary, the unperturbed $SU(6)$ invariant levels
will depend
on the number of quarks and on the way in which the spatial part is
described, that
is an explicit dynamics must be considered.\\
What we have presented here is not the only example of such a situation.
In fact recently in
~\cite{diakonov} a Hamiltonian, containing the
quadratic $SU(3)$ Casimir and a
Gell-Mann Okubo symmetry breaking term, has been used for the calculations
of energy splittings both
for baryons and pentaquarks. This is an indication that different
effective models for the
baryons at the first oder  give origin to a Gell-Mann Okubo mass formula,
independently from the fact
that we consider a Chiral Soliton Model or a bag model or a CQM, that
means independently from which
effective degrees of freedom we use and how we describe from a spatial
point of view the baryon bound
states.

\clearpage
\newpage

\clearpage
\newpage


\begin{table}[!hb]
  \caption{The fitted values of the parameters of the Hamiltonian
(\ref{eq:hcqmhamilt}). 
Column (I) corresponds to the analytical fixing procedure of eq. 
(\ref{eq:fixingparameters}), while column
(II) contains the values obtained with a global fit to the experimental 
resonance masses.}
\label{tab:paramgr1}
    \begin{tabular}{crr}
      \hline\noalign{\smallskip}
      Parameter & (I) & (II) \\ 
      \noalign{\smallskip}\hline\noalign{\smallskip} 
      $\alpha=$ & $1.4$ fm$^{-2}$ & $2.1$ fm$^{-2}$ \\
      $\tau=$   & $-4.8$         & $-3.9$          \\
      $A=$      & $-13.8$ MeV    & $-11.9$ MeV     \\         
      $B=$      & $7.1$ MeV      & $11.7$ MeV      \\
      $C=$      & $38.3$         & $30.8$          \\ 
      $D=$      & $-197.3$ MeV   & $-197.3$ MeV    \\
      $E=$      & $38.5$ MeV     & $38.5$ MeV      \\
      \noalign{\smallskip}\hline
    \end{tabular}
\end{table}


\begin{table*}[ht]
 \caption{Masses of baryon resonances (all values expressed in MeV) calculated
 with the Hamiltonian of eq.(\ref{eq:hcqmhamilt}). 
 The column $M^{(\mbox{\footnotesize I})}_{\mbox{calc}}$ 
 contains the baryon masses calculated with the set of parameters 
 (Table \ref{tab:paramgr1} column (I)) obtained with the 
 analytical fixing procedure (eq. (\ref{eq:fixingparameters})), while the  
 column  $M^{(\mbox{\footnotesize II})}_{\mbox{calc}}$ contains the baryon masses 
 calculated with the parameters of Table \ref{tab:paramgr1} column (II), 
 obtained with a global fit to all the 3 and 4 star experimental 
 data \cite{pdg}.}
\label{tab:tabstrani1}
  \begin{tabular}{lcccrr}
    \hline\noalign{\smallskip}
    Baryon      & Status & Mass      & State & M$^{(I)}_{\mbox{calc}}$ &
    M$^{(II)}_{\mbox{calc}}$\\
    \noalign{\smallskip}\hline\noalign{\smallskip}
    N(938)P11   & ****   & 938       & $^28_{1/2}[56,0^+]$ & 938.0  &
    938.0 \\
    N(1440)P11  & ****   & 1430-1470 & $^28_{1/2}[56,0^+]$ & 1448.7 &
    1455.8 \\ 
    N(1520)D13  & ****   & 1515-1530 & $^28_{3/2}[70,1^-]$ & 1543.7 &
    1492.9\\
    N(1535)S11  & ****   & 1520-1555 & $^28_{1/2}[70,1^-]$ & 1543.7 &
    1492.9\\
    N(1650)S11  & ****   & 1640-1680 & $^48_{1/2}[70,1^-]$ & 1658.6 &
    1585.3\\
    N(1675)D15  & ****   & 1670-1685 & $^48_{5/2}[70,1^-]$ & 1658.6 &
    1585.3\\ 
    N(1680)F15  & ****   & 1670-1685 & $^28_{5/2}[56,2^+]$ & 1651.4 &
    1636.6\\
    N(1700)D13  & ***    & 1650-1750 & $^48_{3/2}[70,1^-]$ & 1658.6 &
    1585.3\\
    N(1710)P11  & ***    & 1680-1740 & $^28_{1/2}[56,0^+]$ & 1795.4 &
    1760.6\\
    N(1720)P13  & ****   & 1650-1750 & $^28_{3/2}[56,2^+]$ & 1651.4 &
    1636.6\\
    $\Delta$(1232)P33 & **** & 1230-1234 & $^410_{3/2}[56,0^+]$ & 1232.0
    & 1251.2\\
    $\Delta$(1600)P33 & ***  & 1550-1700 & $^410_{3/2}[56,0^+]$ & 1683.0
    & 1768.9\\
    $\Delta$(1620)S31 & **** & 1615-1675 & $^210_{1/2}[70,1^-]$ & 1722.8
    & 1713.7\\
    $\Delta$(1700)D33 & **** & 1670-1770 & $^210_{3/2}[70,1^-]$ & 1722.8
    & 1713.7\\
    $\Delta$(1905)F35 & **** & 1870-1920 & $^410_{5/2}[56,2^+]$ & 1945.4
    & 1949.7\\
    $\Delta$(1910)P31 & **** & 1870-1920 & $^410_{1/2}[56,2^+]$ & 1945.4
    & 1949.7\\
    $\Delta$(1920)P33 & ***  & 1900-1970 & $^410_{3/2}[56,0^+]$ & 2089.4
    & 2073.8\\
    $\Delta$(1950)F37 & **** & 1940-1960 & $^410_{7/2}[56,2^+]$ & 1945.4
    & 1949.7\\
    $\Sigma$(1193)P11   & **** & 1193      & $^28_{1/2}[56,0^+] $ &
    1193.0 & 1193.0\\
    $\Sigma$(1660)P11   & ***  & 1630-1690 & $^28_{1/2}[56,0^+] $ &
    1703.7 & 1710.7\\
    $\Sigma$(1670)D13   & **** & 1665-1685 & $^28_{3/2}[70,1^-] $ &
    1798.7 & 1747.9\\
    $\Sigma$(1750)S11   & ***  & 1730-1800 & $^28_{1/2}[70,1^-] $ &
    1798.7 & 1747.9\\
    $\Sigma$(1775)D15   & **** & 1770-1780 & $^48_{5/2}[70,1^-] $ &
    1913.6 & 1840.3\\
    $\Sigma$(1915)F15   & **** & 1900-1950 & $^28_{5/2}[56,2^+] $ &
    1906.4 & 1891.6\\
    $\Sigma$(1940)D13   & ***  & 1900-1950 & $^48_{3/2}[70,1^-] $ &
    1913.6 & 1840.3\\
    $\Sigma^*$(1385)P13 & **** & 1383-1385 & $^410_{3/2}[56,0^+]$ &
    1371.6 & 1390.7\\
    $\Sigma^*$(2030)F17 & **** & 2025-2040 & $^410_{7/2}[56,2^+]$ &
    2085.0 & 2089.2\\
    $\Lambda$(1116)P01  & **** & 1116      & $^28_{1/2}[56,0^+]$  &
    1116.0 & 1116.0\\
    $\Lambda$(1600)P01  & ***  & 1560-1700 & $^28_{1/2}[56,0^+]$  &
    1626.7 & 1633.8\\
    $\Lambda$(1670)S01  & **** & 1660-1680 & $^28_{1/2}[70,1^-]$  &
    1721.7 & 1670.9\\
    $\Lambda$(1690)D03  & **** & 1685-1690 & $^28_{3/2}[70,1^-]$  &
    1721.7 & 1670.9\\
    $\Lambda$(1800)S01  & ***  & 1720-1850 & $^48_{1/2}[70,1^-]$  &
    1836.6 & 1763.3\\
    $\Lambda$(1810)P01  & ***  & 1750-1850 & $^28_{1/2}[56,0^+]$  &
    1973.4 & 1938.6\\
    $\Lambda$(1820)F05  & **** & 1815-1825 & $^28_{5/2}[56,2^+]$  &
    1829.4 & 1814.6\\
    $\Lambda$(1830)D05  & **** & 1810-1830 & $^48_{5/2}[70,1^-]$  &
    1836.6 & 1763.3\\
    $\Lambda$(1890)P03  & **** & 1850-1910 & $^28_{3/2}[56,2^+]$  &
    1829.4 & 1814.6\\
    $\Lambda$(2110)F05  & **** & 2090-2140 & $^28_{5/2}[70,2^+]$  &
    1995.0 & 1957.3\\
    $\Lambda^*$(1405)S01& **** & 1402-1410 & $^21_{1/2}[70,1^-]$  &
    1657.9 & 1565.6\\
    $\Lambda^*$(1520)D01& **** & 1518-1520 & $^21_{3/2}[70,1^-]$  &
    1657.9 & 1565.6\\
    $\Xi$(1318)P11      & **** & 1314-1316 & $^28_{1/2}[56,0^+]$  &
    1332.6 & 1332.5\\
    $\Xi$(1820)D13      & ***  & 1818-1828 & $^28_{3/2}[70,1^-]$  &
    1938.3 & 1887.4\\
    $\Xi^*$(1530)P11    & **** & 1531-1532 & $^410_{3/2}[56,0^+]$ &
    1511.1 & 1530.2\\
    $\Omega$(1672)P03   & **** & 1672-1673 & $^410_{3/2}[56,0^+]$ &
    1650.7 & 1669.7\\
    \noalign{\smallskip}\hline
  \end{tabular}
\end{table*}

\end{document}